# APPROXIMATE K-NEAREST NEIGHBOUR BASED SPATIAL CLUSTERING USING K-D TREE

Dr. Mohammed Otair

Department of Computer Information Systems, Amman Arab University, Amman, Jordan
Otair@aau.edu.jo

## ABSTRACT

*Different spatial objects that vary in their characteristics, such as molecular biology and geography, are presented in spatial areas. Methods to organize, manage, and maintain those objects in a structured manner are required. Data mining raised different techniques to overcome these requirements. There are many major tasks of data mining, but the mostly used task is clustering. Data set within the same cluster share common features that give each cluster its characteristics. In this paper, an implementation of Approximate kNN-based spatial clustering algorithm using the K-d tree is proposed. The major contribution achieved by this research is the use of the k-d tree data structure for spatial clustering, and comparing its performance to the brute-force approach. The results of the work performed in this paper revealed better performance using the k-d tree, compared to the traditional brute-force approach.*

## 1. INTRODUCTION

Spatial data are data that have a location (spatial) characteristic [7] [26]. Spatial data are stored in databases called spatial databases, which contain spatial data type in its data model -side by side- with the ordinary non-spatial types. Spatial data are mainly required for Geographic Information Systems (GIS), whose information is related to geographic locations. GIS model support spatial data types, such as point, line, and polygon [25]. Spatial databases are database systems that manage spatial data. They are designed to manipulate both spatial information and the non-spatial attributes of that data. In order to provide efficient and effective access to spatial data it is necessary to develop indices. These indices are most successful when the data based on multi-dimensional trees. Spatial Database System (SDBS) is a database system designed to manipulate spatial data and the non-spatial data used to recognize the data [8]. The amount of spatial data that be collected in such systems is also increasing exponentially. The complexity of the data found in these databases mean that it is not possible for humans completely to analyze the data being collected. Extremely huge databases require new methods to analyze the data and discover these patterns. Standard database would require additional amounts of space to store spatial data. Moreover, queries to retrieve, analyze spatial data from a these standard database would be too long [8].

Spatial data mining is the operation of applying different mining methods to a spatial database to find non-trivial patterns from the spatial data [7]. One of the most important data mining methods is clustering. Clustering is the task of dividing the objects from spatial database into groups (clusters) in such a way that objects in one cluster are similar and share common features, while objects from different clusters are dissimilar [7] [24]. Clustering helps in discovering and understanding the natural structure and grouping the data in data set. Spatial clustering algorithms must be able to determine clusters of different dimensions, sizes, shapes and density [7] [24].

The k-d tree [14] is a data structure invented by Jon Bentley in 1975. Despite its fairly old age and there exist a number of spatial index structures in literature; however, k-d tree and its variants remain probably the most popular data structures used for searching in multidimensional spaces, at least in main memory. According to The ACM Digital Library, the paper [14] that introduced this data structure is one of the most cited papers in computational geometry, with 626 citations as of February 2013. A k- d tree is multidimensional generalization of a binary search tree. It can be used efficiently for range queries and nearest neighbour searches, provided the dimension is not too high. In document analysis problems, the dimension is typically two, so that k-d trees can be a powerful utility for layout analysis problems. However, K-d trees do not work well in high dimensions (where we have to visit lots and lots of tree branches). Many techniques are developed to overcome these limitations and to perform the search faster [18] [22] [16]. Various improvements to *k*-nearest neighbour methods are possible by using proximity graphs [31].

As mentioned before, one of the earliest data structures proposed for the nearest neighbour problem that is still the most commonly used is the k-d tree [13] that is essentially a hierarchical decomposition of space a long different dimensions. For low dimensions this structure can be used for answering nearest neighbour queries in logarithmic time and linear space. However the performance seems to degrade as a number of dimensions becomes larger than two [20]. For high dimensions, the exact problem of nearest neighbour search seems to suffer from the curse of dimensionality that is either the running time or the space requirement grows exponentially in *d*. Discovering a way to reduce the computational complexity of k-d tree search is of considerable interest in these areas.

This paper will study the problem of finding the approximate k-nearest neighbour based spatial clustering using Kd-tree of a query point in a high dimensional (2D and 3D) space: given a database of *n* points in a *d* dimensional space, find approximate k-nearest neighbour of a query point. The use of k-d trees is a well known optimization to the kNN algorithm [34]. Work on the parallelization of the kNN algorithm in CUDA by Garcia in [10] serves as the starting point for the kNN enhancements is demonstrated. The algorithm they put forward is a parallel implementation of the simple `brute force' kNN algorithm. In the literature they reported a speedup of roughly 100 times faster than a serial CPU implementation and 40 times faster than serial kNN using k-d trees [10]. It was also found that the dimension of points had negligible impact on the overall computation time [11]. One rule of thumb is that if your data dimensionality is $k$, a k-d tree is only going to be any good if it has many more than $2^k$ data points. In high dimensions, it generally wants to switch to approximate nearest-neighbour (ANN) searches instead. ANN is a very useful library for this with C++; it has good implementations of k-d trees, brute-force search, and several approximate techniques, and it helps automatically tune their parameters and switch between them easily.

## 2. RELATED WORK

Most of the current kNN algorithms (such as K-d tree) are too slow for the clustering task. It was determined that the existing K-d tree based nearest neighbour search algorithm suffered performance devaluation from *"curse of dimensionality"* and performance needed improvement [15]. This research will work to proof that the problem can be resolved with an ANN algorithm to speed up the clustering of such speed data. A focus of this research is to improve performance of the KNN approach and to demonstrate its performance in a real-world problem. Another objective of this paper is to test the improvement performance of the existing K-d tree approach.

There are a large number of methods, techniques and algorithms that organize, manage, and maintain spatial objects in a structured manner. However, the most commonly used are:

## 2.1. K-Nearest Neighbour

T. M. Cover and P. E. Hart purpose k-nearest neighbour (kNN) in which nearest neighbour is calculated on the basis of value of *k*, that specifies how many nearest neighbours are to be considered to define class of a sample data point [30]. T. Bailey and A. K. Jain improve kNN which is based on weights [28]. The training points are assigned weights according to their distances from sample data point. But still, the computational complexity and memory requirements remain the main concern always [18]. To overcome memory limitation, size of data set should be reduced.

The k-nearest neighbour join combines each point of one point set with its *k* nearest neighbours [5]. The general model of a KNN query is that the user gives many query types such a point query in multidimensional space and a distance metric for measuring distances between points in this space. The system is then tried to find the *K* closest or nearest answers in the database from the submitted query (i.e. query point). Generally distance metrics may include: Euclidean distance, Manhattan distance, etc. It is possible that a majority of the answers to a KNN query may be very similar to one or more of the other answers, especially when the data has clusters [1]. Given a set of *n* points in a d-dimensional space, the k-d tree is constructed recursively as follows. First, one finds a median of the values of the *ith* coordinates of the points (initially, $i = 1$). That is, a value *M* is computed, so that at least 50% of the points have their *ith* coordinate greater-or-equal to M, while at least 50% of the points have their *ith* coordinate smaller than or equal to *M*. The value of *x* is stored, and the set *P* is partitioned into *PL* and *PR*, where *PL* contains only the points with their *ith* coordinate smaller than or equal to *M*, and |PR| = |PL| ±1. The process is then repeated recursively on both *PL* and *PR*, with *i* replaced by *i + 1* (or 1, if i = d). When the set of points at a node has size 1, the recursion stops. The kNN implementation can be done using box-decomposition trees (ball tree) [23] [29], k-d tree [19], these algorithms increase the speed of basic kNN algorithm. The k-Nearest Neighbour (kNN) algorithm has a wide range of applications in modern day computing.

The simplest version of the kNN algorithm is the 'Brute Force' implementation and consists of three stages. The first stage is to calculate all of the `distances' from each query point to every reference point in the training set. The second stage is to sort these distances and select the *k* objects that are the closest from which the third and final stage of classification can be performed. More formally, KNN finds the *K* closest (or most similar) points to a query point among *N* points in a d- dimensional attribute (or feature) space. *K* is the number of neighbours that are considered from a training data set and typically ranges from 1 to 20. The **advantages** of k Nearest Neighbour (kNN) can be summarized as in [30]: Training is very fast, Simple and easy to learn, Robust to noisy training data, and Effective if training data is large. However, there many disadvantages**,** as well: Biased by value of *k,* Computation Complexity, Memory limitation, being a supervised learning lazy algorithm i.e. runs slowly, and easily fooled by irrelevant Attributes

## 2.2. Spatial Indexing

In order to manipulate spatial data efficiently, as required in some geo-data applications, which are concerned with spatial search on multi-dimensional spaces, a database system needs an index mechanism that will help it in retrieving and accessing data items quickly according to their spatial locations. However, at the other side traditional indexing methods are not well suited to data objects of non-zero size located in multi-dimensional spaces [2].

The most important requirements for these data structures are the ability to provide fast access to large volumes of data and preserve spatial relations such as nesting and neighbourhood for indexed objects.

Several tree-like access methods were proposed for spatial objects [33] such as R-tree which is a hierarchical data structure. It is for efficient indexing of multidimensional objects with spatial extent [2] [33]. Several other tree index structures exist in the literature such as the R+-tree, R*-tree [17] [33] and many other tree index structures.

In this paper, the KD-tree will be used for spatial indexing in a new KNN-based spatial clustering algorithm.

### 2.3. K-d Tree

A k-d tree, or k-dimensional tree, is a data structure used for organizing some number of points in a space with k dimensions. It is a binary search tree with other constraints imposed on it. K-d trees are very useful for range and nearest neighbour searches. The *root-cell* of this tree represents the entire simulation volume. The other cells represent rectangular sub-volumes that contain the mass, center-of-mass, and quadrupole moment of their enclosed regions.

It was one of the early structures used for indexing in multiple dimensions. Each level of K-d tree partitions the space into two partitions, as shown in figure 2.1, the partitioning is done along one dimension of the node at the top level of the tree, along another dimension in nodes at the next level, and so on, iterating through the dimensions. The partitioning proceeds in such a way that, at each node, approximately one half of the points stored in the subtree fall on one side, and one half fall on the other. Partitioning stops when a node has less than a given maximum number of points [32].

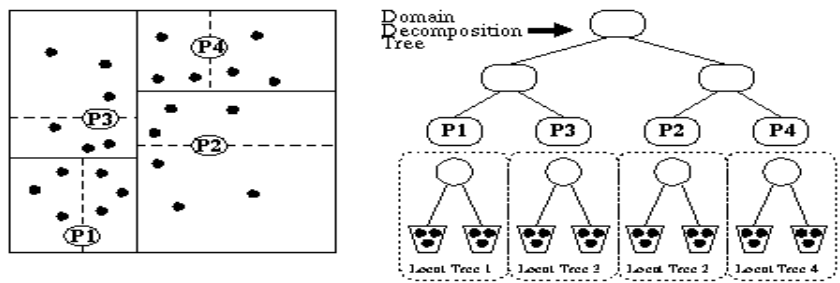

**Figure 2.1:** K-d Tree Partitioning

Although many different flavours of k-d trees have been devised, their purpose is always to hierarchically decompose space into a relatively small number of cells such that no cell contains too many input objects. This provides a fast way to access any input object by position. We traverse down the hierarchy until we find the cell containing the object. Typical algorithms construct k-d trees by partitioning point sets recursively along with different dimensions. Each node in the tree is defined by a plane through one of the dimensions that partitions the set of points into left/right (or up/down) sets, each with half the points of the parent node. These children are again partitioned into equal halves, using planes through a different dimension. Partitioning stops after log n levels, with each point in its own leaf cell. The partitioning loops through the different dimensions for the different levels of the tree, using the median point for the partition. K-d trees are known to work well in low dimensions but seem to fail as the number of dimensions increase beyond three [27].

K-d tree nearest neighbour (kdNN) [19] has many advantages such as: Produce perfectly balanced tree, and Fast and simple. However, it has some important disadvantages**:** More computation, require intensive search, and blindly slice points into half which may miss data structure

## 2.4. Spatial Clustering

A lot of work has been done in the area of clustering. Several clustering algorithms and techniques have been developed and implemented. The most common are:

### 2.4.1. Partitioning algorithms

Partitioning algorithms construct a partition of a database of many objects into a set of clusters (k clusters), it constructs the clusters in one step as opposed to several steps, and only one set of clusters is constructed, although several different sets of clusters may be constructed internally within the various algorithms. Since only one set of clusters is output, the user must input the desired number of clusters. Each cluster is represented by one of the objects of the cluster located near its center (k-medoid algorithms) or by the center of the cluster (k-means algorithms) [12].

### 2.4.2. Hierarchical algorithms

Hierarchical clustering organizes clusters into a tree or a hierarchy structure represented using a tree data structure called a dendrogram, where the root of this tree is a single cluster that contains all objects in the spatial area. Leaves are clusters with single object. Hierarchical algorithms run iteratively to perform splitting or merging until a stopping criterion is met or all objects have been clustered. Hierarchical algorithms are categorized into agglomerative and divisive algorithms, where the dendrogram can either be created from the leaves up to the root (Agglomerative approach), or from root down to leaves (divisive approach).

### 2.4.3. Density-based algorithms

These algorithms are based on density, such as density-connected points. Such algorithms have many features, such as its ability to discover clusters of arbitrary shapes, as well as, to distinguish noise, and they require a density parameter as a termination condition [12].

### 2.4.4. Grid-based algorithms

Grid-based algorithms depend on the size of the grid instead of the data objects, using a single uniform grid mesh to partition the entire problem domain into cells, and the data objects within a cell are represented by the cell using a set of statistical attributes from the objects. Grid-based algorithms perform clustering on the grid cells, instead of the database itself. Therefore, these algorithms have the fastest processing time [12].

The run time of most of the above algorithms is too inefficient on large databases. Therefore, some focusing techniques have been proposed to improve the efficiency of clustering algorithms. The researchers in [9] discussed the use of an R*-tree in spatial clustering. Moreover, based-focusing techniques were introduced in [4] [9].  These focusing techniques proceed in the following phases: (1) creating a sample of the database that is drawn from each R*-tree data page and (2) applying the clustering algorithm only to that sample.

This paper introduces a new KNN- based spatial clustering algorithm using the Kd-tree, which will be described in the following section.

## 3. APPROXIMATE K-NEAREST NEIGHBOUR BASED SPATIAL CLUSTERING

This paper is concerned with the problem of Approximate KNN based spatial clustering. The concept is based on clustering spatial points that are the most nearest and have similar properties into one cluster. In order to compute the nearest neighbours, a simple brute-force search can be used. However, in order to handle large volumes of spatial data organized in high dimensions, brute force will be too slow. Therefore, the need arises for other techniques that are based on indexing spatial data into a data structure that can be used to answer nearest neighbours queries. In this paper, the k-d tree is chosen as the data structure to index spatial points. In order to evaluate the efficiency of the k-d tree for achieving this purpose, several experiments have been performed to evaluate and compare the efficiency of the proposed method when applied on various data size, various dimensions, and multiple *k* values. Implement approximate *k-nearest-neighbour (kNN)* search using a brute force approach as well as with the help of the kd-tree will be used to reach of the main objective of this research (i.e. to speed-up K-nearest neighbour searches). Then, the obtained results will be presented, discussed, and analyzed in the next section.

### 3.1. Brute Force Algorithm

Brute-force search or exhaustive search is a very general problem-solving technique that consists of systematically enumerating all possible candidates for the solution and checking whether each candidate satisfies the problem's statement. The simple sequence of operations for the brute force algorithm is shown below:
1. Compute all the distances between the query point and reference points
2. Sort the computed distances.
3. Select the k reference points with the smallest distances
4. Classification vote by k nearest objects.
5. Repeat steps (1 to 4) for all query points.

### 3.2. K-d Tree Algorithm

The *k*-d tree is a binary tree in which every node is a k-dimensional point. Every non-leaf node can be thought of as implicitly generating a splitting hyper-plane that divides the space into two parts, known as half-spaces. Points to the left of this hyper-plane represent the left subtree of that node and points right of the hyper-plane are represented by the right subtree. The hyper-plane direction is chosen in the following way: every node in the tree is associated with one of the k-dimensions, with the hyper-plane perpendicular to that dimension's axis. So, for example, if for a particular split the "x" axis is chosen, all points in the subtree with a smaller "x" value than the node will appear in the left subtree and all points with larger "x" value will be in the right subtree. In such a case, the hyper-plane would be set by the x-value of the point, and its normal would be the unit x-axis [14].

The nearest neighbor search (NN) algorithm aims to find the point in the tree that is nearest to a given input point. This search can be done efficiently by using the tree properties to quickly eliminate large portions of the search space. Searching for a nearest neighbor in a *k*-d tree proceeds as follows [Source: Wikipedia]:

1. Starting with the root node, the algorithm moves down the tree recursively, in the same way that it would if the search point were being inserted (i.e. it goes left or right depending on whether the point is less than or greater than the current node in the split dimension).
2. Once the algorithm reaches a leaf node, it saves that node point as the "current best"
3. The algorithm unwinds the recursion of the tree, performing the following steps at each node:

   1. If the current node is closer than the current best, then it becomes the current best.
   2. The algorithm checks whether there could be any points on the other side of the splitting plane that are closer to the search point than the current best. In concept, this is done by intersecting the splitting hyper-plane with a hyper-sphere around the search point that has a radius equal to the current nearest distance. Since the hyper-planes are all axis-aligned this is implemented as a simple comparison to see whether the difference between the splitting coordinate of the search point and current node is less than the distance (overall coordinates) from the search point to the current best.
      1. If the hyper-sphere crosses the plane, there could be nearer points on the other side of the plane, so the algorithm must move down the other branch of the tree from the current node looking for closer points, following the same recursive process as the entire search.
      2. If the hyper-sphere doesn't intersect the splitting plane, then the algorithm continues walking up the tree, and the entire branch on the other side of that node is eliminated.
   4. When the algorithm finishes this process for the root node, then the search is complete.

Generally the algorithm uses squared distances for comparison to avoid computing square roots. Additionally, it can save computation by holding the squared current best distance in a variable for comparison.

### 3.3. Approximate Nearest Neighbour (ANN)

In some applications it may be acceptable to retrieve a "good guess" of the nearest neighbour. In those cases, we can use an algorithm which doesn't guarantee to return the actual nearest neighbour in every case, in return for improved speed or memory savings. Often such an algorithm will find the nearest neighbour in a majority of cases, but this depends strongly on the dataset being queried. Algorithms that support the approximate nearest neighbour search include locality-sensitive hashing, best bin first and balanced box-decomposition tree based search [21]

The k-d tree is searched for an approximate nearest neighbour. The point is returned through one of the arguments, and the distance returned is the squared distance to this point. The method used for searching the k-d tree is an approximate adaptation of the search algorithm described by Friedman, Bentley, and Finkel in [13].

The algorithm operates recursively. When first encountering a node of the k-d tree we first visit the child which is closest to the query point. On return, we decide whether we want to visit the other child. If the box containing the other child exceeds 1/(1+eps) times the current best distance, then we skip it (since any point found in this child cannot be closer to the query point by more than this factor.) Otherwise, we visit it recursively. The distance between a box and the query point is computed exactly not approximated as is often done in k-d tree), using incremental distance updates, as described by Arya and Mount in [3].

The main entry points to the ANN search sets things up and then call the recursive routine another search routine. This is a recursive routine which performs the processing for one node in the k-d tree. There are two versions of this virtual procedure, one for splitting nodes and one for leaves. When a splitting node is visited, we determine which child to visit first (the closer one), and visit the other child on return. When a leaf is visited, we compute the distances to the points in the buckets, and update information on the closest points. Some trickery is used to incrementally update the distance from a k-d tree rectangle to the query point. This comes about from the fact that which each successive split, only one component along the dimension that is split) of the squared distance to the child rectangle is different from the squared distance to the parent rectangle.

ANN is a library written in the C++ programming language to support both exact and approximate nearest neighbour searching in spaces of various dimensions. It was implemented by David M. Mount of the University of Maryland and Sunil Arya of the Hong Kong University of Science and Technology.  ANN (pronounced like the name \Ann") stands for the Approximate Nearest Neighbour library. ANN is also a testbed containing programs and procedures for generating data sets, collecting and analyzing statistics on the performance of nearest neighbour algorithms and data structures, and visualizing the geometric structure of these data structures.

Answering nearest neighbour queries efficiently, especially in higher dimensions, seems to be very difficult problem. It is always possible to answer any query by a simple brute-force process of computing the distances between the query point and each of the data points, but this may be too slow for many applications that require that a large number of queries be answered on the same data set. Instead the approach is to pre-process a set of data points into a data structure from which nearest neighbour queries are then answered. There are a number of data structures that have been proposed for solving this problem [6].

One difficulty with exact nearest neighbour searching is that for virtually all methods other than brute-force search, the running time or space grows exponentially as a function of dimension. Consequently these methods are often not significantly better than brute-force search, except in fairly small dimensions. However, it has been shown by Arya and Mount and Arya, et al. that if the user is willing to tolerate a small amount of error in the search (returning a point that may not be the nearest neighbour, but is not significantly further away from the query point than the true nearest neighbour) then it is possible to achieve significant improvements in running time. ANN is a system for answering nearest neighbour queries both exactly and approximately [6].

ANN was written as a testbed for a class of nearest neighbour searching algorithms, particularly those based on orthogonal decompositions of space. These include k-d trees, and box-decomposition trees. The library supports a number of different methods for building these search structures. It also supports two methods for searching these structures [6]: standard tree-ordered search and priority search.

## 4. EXPERIMENTAL EVALUATION AND RESULTS ANALYSIS

The Approximate KNN–based spatial clustering method has been tested through several experiments using the k-d tree. The nearest neighbour spatial distance was computed using the ANN library, which was modified to perform spatial clustering. The experiments have been performed on a 2.8 GHz PC, with Intel Pentium 4 processor and 256 MB memory. The ANN library has been compiled under visual C++ .Net 2003.

The experiments were conducted for computing 1, 2, 3, 4 and 5 nearest neighbours, with data sets of sizes 0.5 and 1 MB assuming the spatial points are organized in 2 and 3 dimensions.

The first set of experiments has been performed to compare the performance of both, K-d tree and brute-force, for a 0.5 MB data set in a 2-dimentional space. The execution time required by both approaches was approximately the same. Better performance was achieved when the k-d tree was used for larger number of K nearest neighbours. Figure 4.1 shows the execution time required by K-d tree when 1, 2, 3 and 4 nearest neighbours are computed. However, for 5 nearest neighbours, better performance was observed using the K-d tree.

104

International Journal of Database Management Systems ( IJDMS ) Vol.5, No.1, February 2013

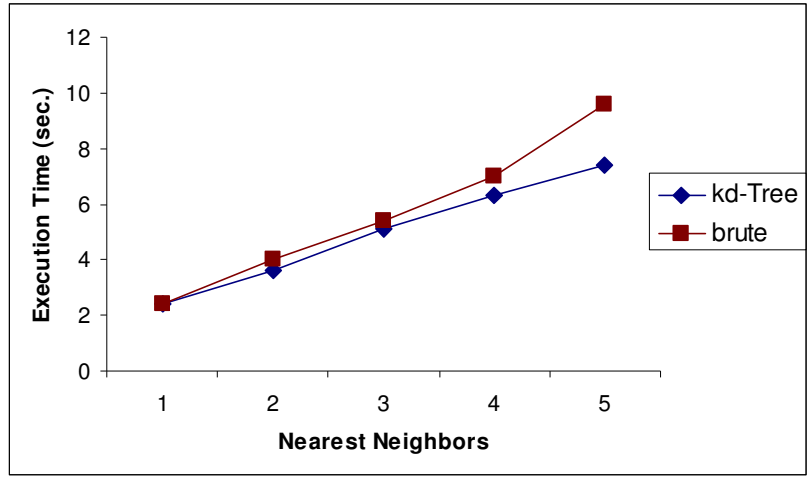

**Figure 4.1: K-d Tree vs. brute-force Performance (0.5 MB data set, 2-D)**

The next set of experiments has been conducted for the same data set size (0.5 MB), however; in a 3-dimensional space. It is obvious from figure 4.2 that as the number of nearest neighbours increases; the K-d tree shows better performance.

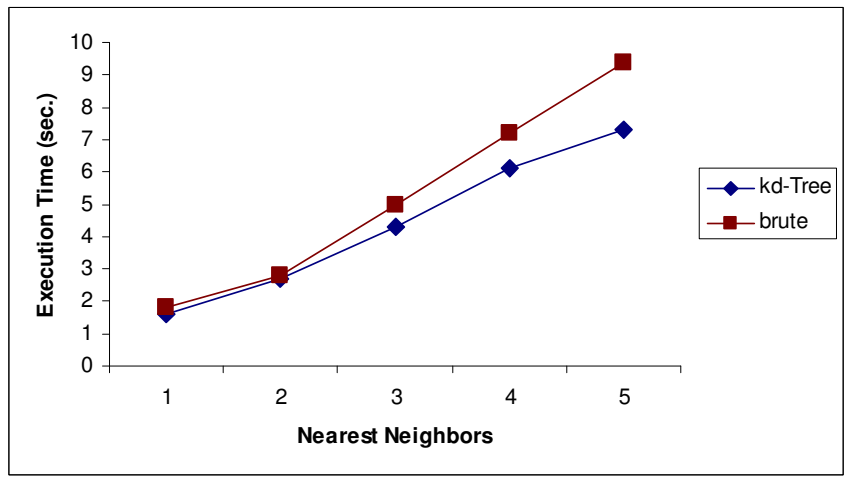

**Figure 4.2: K-d Tree vs. brute-force Performance (0.5 MB data set, 3-D)**

With larger volumes of data, an enhanced performance was achieved by K-d tree. The enhancement becomes clearer for data organized in 3 dimensions. These results are clear from figures 4.3 and 4.4.

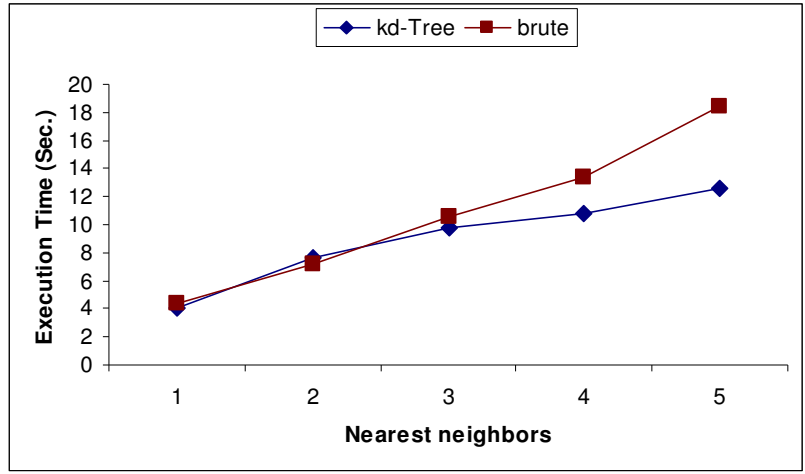

**Figure 4.3: K-d Tree vs. brute-force Performance (1 MB data set, 2-D)**

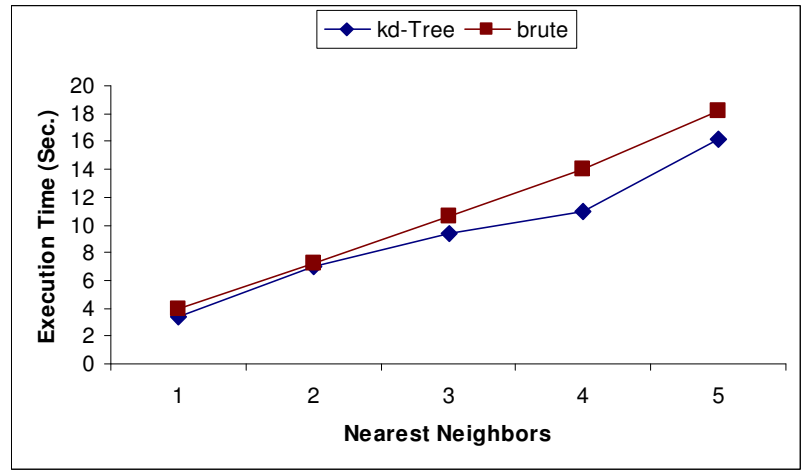

**Figure 4.4: K-d Tree vs. brute-force Performance (1 MB data set, 3-D)**

As seen in the previous figures (4.3 and 4.4), the K-d Tree algorithm needs less number of seconds in comparing with brute-force especially when using a large volumes of data regardless the number of dimensions (i.e. if it is in 2D or 3D).

## 5. CONCLUSIONS

The major contribution achieved by this research is the use of the approximate k-nearest neighbour with k-d Tree data structure for spatial clustering, and comparing its performance to the brute-force approach. The results of the work performed in this paper revealed better performance using the k-d Tree, compared to the traditional brute-force approach. The efficiency of the data structure primarily depends on a particular implementation and data set. A poorly balanced tree will mean we have to search way more data than we need to.

## FUTURE WORK

As a future work, other data structures can be used to achieve spatial clustering. The results obtained can be compared to this paper's results.

## ACKNOWLEDGMENT

The Author would like to thank Sama Al-Momani and Zeinab Jaradat (My Students) for their help in the experiments that done in this research.

**Author**

**Mohammed A. Otair** is an Associate Professor in Computer Information Sy University-Jordan. He received his B.Sc. in Computer Science from IU-Jordan an 2000, 2004, respectively, from the Department of Computer Information Syste major interests are Mobile Computing, Data Mining and Databases Neural Netwo Web-computing, E-Learning. He has more than 29 publications.

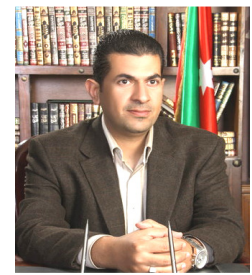